*Letter to the Editor: first results from ISO*

# The Mid-Infrared Color of NGC 6946*

G. Helou[1], S. Malhotra,[1], C. A. Beichman[1], H. Dinerstein[2], D. J. Hollenbach[3], D. A. Hunter[4], K. Y. Lo[5], S. D. Lord[1], N. Y. Lu[1], R. H. Rubin[3], G. J. Stacey[6], H. A. Thronson Jr.[7], and M. W. Werner[8]

[1] IPAC, California Institute of Technology, MS 100-22, Pasadena, CA 91125
[2] University of Texas, Astronomy Department, RLM 15.308, Texas, Austin, TX 78712
[3] NASA/Ames Research Center, MS 245-6, Moffett Field, CA 94035
[4] Lowell Observatory, 1400 Mars Hill Rd., Flagstaff, AZ 86001
[5] University of Illinois, Astronomy Department, 1002 W. Green St., Urbana, IL 61801
[6] Cornell University, Astronomy Department, 220 Space Science Building, Ithaca, NY 14853
[7] University of Wyoming, Wyoming Infrared Observatory, Laramie, WY, 82071
[8] Jet Propulsion Laboratory, MS 233-303, 4800 Oak Grove Rd., Pasadena, CA 91109

**Abstract.** We analyze the new mid-infrared maps of NGC 6946 for variations in the color ratio of the 7-to-15$\mu$m emission. Our preliminary findings are that this mid-infrared color is remarkably constant between arms and inter-arm regions, and as a function of radius in the disk, excluding the nuclear region. As surface brightness ranges by more than an order of magnitude and the radius runs from about 0.5 to 3 kpc, the color ratio remains constant to about $\pm 20\%$. Our interpretation is that (1) hard UV radiation from OB stars does not dominate the heating of the grains radiating in the mid-infrared; and (2) that surface brightness variations are driven primarily by surface-filling fraction in the disk, and by radiation intensity increases in starburst environments, such as the nucleus of NGC 6946.

## 1. Introduction

The Infrared Space Observatory (ISO; Kessler et al. 1996) provides a unique opportunity for studying star formation in systems inaccessible to sub-orbital platforms in infrared (IR) spectroscopy and low-brightness imaging. The "ISO Key Project" on the interstellar medium of normal galaxies, carried out under the NASA Guaranteed Time, uses ISO to derive the physical properties of the interstellar gas, dust and radiation field in a broad sample of "normal" disk galaxies, defined as systems whose luminosity is dominated by star formation. The project collects a variety of diagnostics from ISO instruments, mainly ionic and atomic fine-structure line fluxes using LWS (Clegg et al. 1996), far-IR continuum fluxes and mid-IR spectra using PHOT-C and PHOT-S (Lemke et al. 1996), and mid-IR maps at 7 and 15$\mu$m using CAM (C. Césarsky et al. 1996). Two groups of objects are included: About six nearby galaxies such as NGC 6946 provide spatially resolved cases where the various phases of the interstellar medium (ISM) can be studied separately, and their signatures in each observable identified. About sixty distant galaxies with small apparent IR sizes were selected to span the full ranges of morphology, luminosity, IR-to-blue ratio, and IRAS colors that are covered by star-forming galaxies. By characterizing the variation in ISM properties in this diverse sample, we hope to gain new insight into the the star formation process on the scale of galaxies, especially its drivers and inhibitors.

We report in this paper on analysis of the first data obtained for the Key Project on normal galaxies, namely ISO-CAM maps at 7 and 15$\mu$m of NGC 6946 (Malhotra et al. 1996). The aim is to understand the mid-IR colors and their variation within the disk of this spiral galaxy. IR colors formed among the IRAS bands are well known as sensitive indices of the radiation intensity in the ISM of galaxies (Soifer, Houck & Neugebauer 1987; Helou & Wang 1996).

Send offprint requests to: helou@ipac.caltech.edu
* Based on observations with ISO, an ESA project with instruments funded by ESA Member States (especially the PI countries: France, Germany, the Netherlands and the United Kingdom) and with the participation of ISAS and NASA.

IRAS data established that mid-IR ($5 \leq \lambda \leq 40\mu m$) emission from the ISM and star-forming galaxies is dominated by fluctuating grains and Polycyclic Aromatic Hydrocarbons (PAH) whereas classical grains in thermal equilibrium dominate at longer wavelengths (Helou 1986). Many questions however remain about the precise nature of the PAH (Puget & Léger 1989), their life cycle and excitation conditions, some of which we address by a detailed study of the mid-IR colors of galaxies.

Early results from ISO already show the $7\mu m$ filter of CAM (LW2, $\Delta\lambda = 3.5\mu m$) to be primarily sensitive to major PAH emission features at 6.3 and $7.7\mu m$, whereas the 15 $\mu m$ filter (LW3, $\Delta\lambda = 6\mu m$) picks up emission from a minor feature at $12.6\mu m$, but mostly from any continuum at wavelengths up to $\sim 18\mu m$ due to fluctuating very small grains (Vigroux et al. 1996; D. Césarsky et al. 1996). The 7-to-$15\mu m$ ratio should thus measure a feature to continuum ratio, and might be expected to gauge the PAH to very small grain ratio.

## 2. Observations and Data reduction

NGC 6946 was mapped at $7\mu m$ and at 15 $\mu m$ using the raster scan mode to cover roughly $12.5' \times 12.5'$ centered on the nucleus. CAM was set to $6''$/pixel, and the raster was made up of $8 \times 8$ pointings separated by $81''$, or 13.5 pixels, in each direction for better spatial sampling. The details of data reduction and image reconstruction are given in the companion paper by Malhotra et al. (1996). These maps are based on a preliminary data reduction and overall calibration with a 30% uncertainty, so we will not address here the absolute value of the color ratios.

The images resulting from this preliminary data reduction were constructed on a grid with $3'' \times 3''$ pixels, and achieved a spatial resolution with a FWHM of $\sim 7.2''$ at both wavelengths. The noise levels in the images are approximately $0.1\,\mathrm{MJy\,sr}^{-1} \simeq 2\mu\mathrm{Jy\,arcsec}^{-2}$. For comparison, the cleaner images in IRAS Sky Survey Atlas (Wheelock et al. 1994) reach a noise level about three times lower at $12\mu m$ in a $4'$ beam, i.e. a solid angle 1600 times greater than the ISO-CAM images. Features roughly a thousand times brighter than the noise are still reliably measured in these maps. The inner $15''$ or so are too bright to measure, with the surface brightness at the nucleus exceeding 300 and $560\,\mathrm{MJy\,sr}^{-1}$ respectively at 7 and $15\mu m$.

Mid-IR color estimation, especially in the outer regions of the galaxy, is very sensitive to the subtraction of the foreground zodiacal light. Zodiacal light was determined from an annulus of radius $5'$ centered on the nucleus of NGC 6946 after de-projection of the map assuming an inclination angle of $30°$ and a major axis position angle of $67°$ measured East of North. At $7\mu m$ the zodiacal surface brightness equals the galaxy's brightness at a radius of $2'$, and is $\sim 80$ times the noise level in the final map. Spatial gain variations in the maps are another error source, since zodiacal light and galaxy have very different mid-IR colors, with the latter showing a 7-to-$15\mu m$ ratio 4 to 5 times greater.

## 3. Color Gradient in the Disk

The ISOCAM maps of NGC 6946 have enough resolution, sensitivity and extent to allow a characterization of the mid-IR color behavior as a function of position in the disk. Mid-IR morphology at both 7 and $15\mu m$ is generally similar to optical, radio and $H\alpha$ images, with an exponential disk and well defined though flocculent arms (Malhotra et al. 1996). Diffuse emission from the disk in the interarm regions is clearly detected and can be traced out to almost $5'$. The disk is relatively symmetric, making it unlikely to be the result of transient effects in detectors or other artifacts. A simple ratio map of the 7 to $15\mu m$ images shows essentially no sign of structure, suggesting that the colors are relatively constant across disk and arms.

In their analysis of scale lengths, Malhotra et al. (1996) show the radial profile of the median disk brightness to be indistinguishable between the maps at 7 and $15\mu m$. This is compelling evidence for the absence of any radial gradient in the "mean" mid-IR color even as the median brightness drops by a factor of $\sim 10$.

## 4. Arm-Interarm Color Contrast

In order to address the question of color variation between arm and inter-arm, we introduce a relatively objective definition of physically similar and contiguous regions based on brightness. Having deprojected the image of the galaxy to face-on, we consider the distribution of surface brightness in all pixels within annuli $15''$ wide, and take the arms to be those pixels that were the top-ranked 25% in both the $7\mu m$ and $15\mu m$ distributions. Similarly, the lowest-ranked 25% of pixels are taken to be the diffuse disk or inter-arm regions. The result of that sorting is shown in Figure 1, which demonstrates that the method identifies reasonably well arms that would be picked out by eye. The transition region around the median of the distribution is not considered further in the analysis. The two-map condition on the brightness selection is necessary to avoid biasing the color ratio by selecting pixels that are *a priori* high in either the numerator or denominator.

The colors are then estimated for each of the arm or interarm regions, and in each $15''$ annulus, as the ratio between the integrated surface brightnesses at 7 and $15\mu m$. The run of 7-to-$15\mu m$ colors is plotted against radius in Figure 2, and against surface brightness in Figure 3. The uncertainty bars shown reflect two extremes of possible errors, namely that the zero-point subtraction is off by one $\sigma$ in opposite directions for each of the numerator and denominator in the color ratio, where $\sigma$ is the noise in the empty part of the map, $\sim 0.1\,\mathrm{MJy\,sr}^{-1}$. This is clearly a conservative estimate of the possible error in the zero-point determination.

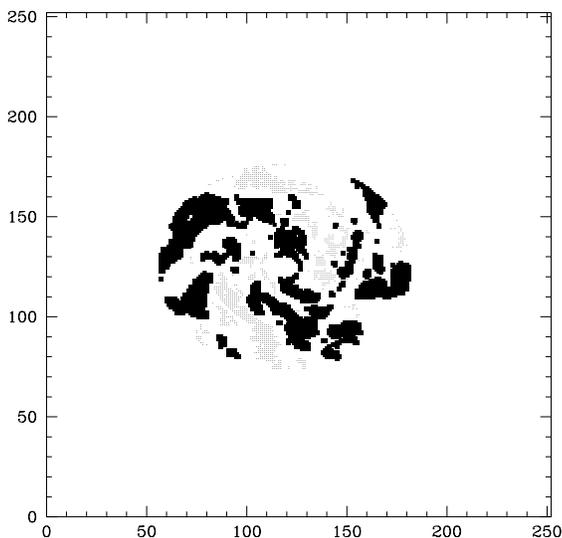

**Fig. 1.** Selection map showing the areas assigned to arm and inter-arm by the mid-IR brightness sorting method described in §3. The dark areas correspond to arms, whereas the lightly shaded areas correspond to diffuse disk or inter-arm regions, and blank areas to the transition regions which are not considered further here. The axes are labelled in number of $3 \times 3''$ pixels.

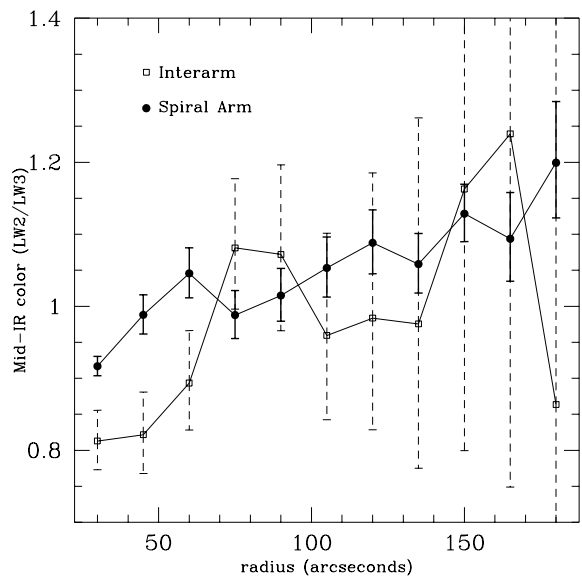

**Fig. 2.** Variation of the 7-to-15$\mu$m color ratio as a function of radius for each of the arms and inter-arm regions. Note that both here and in Figure 3 we have excluded the nuclear region where the galaxy's brightness exceeds the linearity range of the detector. The uncertainty bars for the colors in the arms are shown as solid lines, and those in the inter-arm regions as broken lines.

It is clear from Figure 2 that the mid-IR colors are generally quite constant across the disk. There is a slight trend for an increase in the 7-to-15$\mu$m ratio with increasing radius, but this trend is not significant in view of the uncertainties assumed. This analysis complements the results discussed in §3, since those results pertain to the median brightness whereas these plots address the upper and lower quartiles in the brightness distribution. We conclude that the 7-to-15$\mu$m color is constant to within $\pm 20\%$ in the disk of NGC 6946. This conclusion applies whether one compares arms and diffuse disk, or whether one considers the brighter, medium or dimmer parts of the disk as a function of radius.

In the innermost three annuli plotted in Figure 1, arms and inter-arm regions show slightly different colors, with a significance on the order of 2$\sigma$ only. Given the preliminary status of the data reduction we take a cautious view of the reality of this difference, especially that the detectors would be most susceptible to systematic problems here because of the very bright nuclear region.

Figure 3 shows no sytematic dependence of color on surface brightness, which would have been a direct indication of errors in the zodiacal foreground subtraction or of non-linearities in the flux-scale calibration of either one of the two CAM maps. It again illustrates the basic constancy of color as the surface brightness ranges by a factor of 20. Interestingly, a color difference similar to the 2$\sigma$ effect suggested by Figure 2 appears more significant in Figure 3. Where the brightest inter-arm regions and the dimmest arm segments have comparable brightness around 7 MJy sr$^{-1}$, the arms show a 7-to-15$\mu$m ratio greater than in the inter-arms by almost 50% at the 4$\sigma$ significance level.

The CAM data are consistent with the 7-to-15$\mu$m color dropping further as one approaches the nucleus, but are not sufficiently reliable to draw firm conclusions (Malhotra et al. 1996). Such a change in color however would be consistent with the data in the Antennae galaxies (Vigroux et al. 1996), where the 7-to-15$\mu$m ratio drops by a factor of 2 in the very bright "interaction region" where a star-burst is in progress just as one is known to be occurring in the nucleus of NGC 6946.

## 5. Discussion

The constancy of mid-IR color across the disk and against variations in the surface brightness is somewhat surprising, since one would expect the heating spectrum to change substantially between arm and inter-arm regions, as the former contains more HII regions, massive stars, and therefore a heating spectrum richer in UV photons. The lack of variations suggests that the spectral shape of the emission between 5 and 18$\mu$m is not sensitive at the 20% level to the heating spectrum. By contrast, the combination of PAH and very small grain emission, gauged by the IRAS 12-to-25$\mu$m ratio, has been shown to vary with heating intensity and presumably UV-richness of the heating spectrum (e.g. Boulanger et al. 1988). While it may be argued that the 5 to 18$\mu$m spectral range covered here is too small to show the effects observed in the IRAS colors, it does remain puzzling that the heating spectrum affects so little the feature-to-continuum ratio. One implication is that the relatively soft spectrum in the diffuse regions is adequate to excite emission in the 5 to 18$\mu$m range. This agrees with the Malhotra et al. (1996) observation that the mid-IR contrast between arms and inter-arms is closer to the contrast in the visible R band than to that in the H$\alpha$, implying that ionizing and hard non-ionizing UV radiation from OB stars does not dominate the heating of the grain populations radiating in the mid-IR.

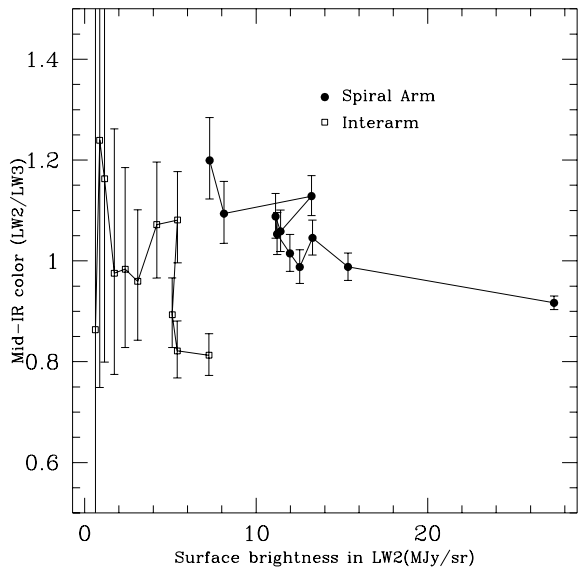

**Fig. 3.** Variation of the 7-to-15$\mu$m color ratio as a function of 7$\mu$m surface brightness for the arms and inter-arm regions. The lines connecting the symbols indicate the radial ordering of the annuli from which the plotted data are derived.

One could speculate however that the difference between the two color sequences in Figure 3, despite the limited significance, does illustrate the distinction between the emission spectra in UV-rich regions and those from diffusely heated regions. Figure 3 would then suggest that UV-rich heating leads to larger ratios of mid-IR PAH features to continuum emission from very small grains. This would agree with the models of Puget & Léger (1989) who attribute greater UV cross-sections to the PAHs. The true enhancement in feature-to-continuum ratio however cannot be estimated from Figure 3 until one knows which pairs of points from the arm and inter-arm color sequences have the same dust column density. Even with such a correspondence established, the ratio variations would have to be corrected for potential PAH destruction in UV-rich environments (Helou et al. 1991).

The shift in mid-IR color observed in an extreme starburst environment (Vigroux et al. 1996) suggests that even the restricted 5 to 18$\mu$m range does eventually get affected by great increases in heating intensity or hardness of heating spectrum, presumably when grains in thermal equilibrium exceed $\sim$100 K and start contributing substantially to the 15$\mu$m band. The fact that this is not approached gradually as the surface brightness rises in the disk and arms (and if anything the reverse is true at 7 MJy sr$^{-1}$ in Figure 3) suggests that the increase in surface brightness does not reflect a heating intensity increase in the disk, but rather a surface filling factor increase. Presumably the arms are brighter because of a greater surface density of distinct and independent star forming regions, each of which emits a similar spectrum in the mid-IR. As star-burst conditions are reached however, the density of HII regions increases to the point of superposition and the heating intensity increases dramatically, causing the surface brightness to increase by at least another order of magnitude.

This analysis leads us to predict that most galaxies will display 7-to-15$\mu$m colors similar to the ones observed in NGC 6946, with any exceptions signalling starburst conditions, meaning extremely high heating intensities and UV-rich heating spectra. Spatially unresolved galaxies with both normal and star-burst components will of course display intermediate mid-IR colors.

*Acknowledgements.* This work was supported by ISO data analysis funding from the US National Aeronautics and Space Administration, and carried out at the Infrared Processing and Analysis Center and the Jet Propulsion Laboratory of the Califronia Institute of Technology. We are particularly grateful to Dave Van Buren and Mihseh Kong for their work with the data reduction.